\documentclass[10pt]{article}
\usepackage[latin1]{inputenc}
\usepackage[T1]{fontenc}
\usepackage{amsmath}
\usepackage{amsfonts}
\usepackage{amssymb}
\usepackage{graphicx}
\usepackage{color}
\usepackage{url}
\usepackage{numprint}
\usepackage{authblk}
\npdecimalsign{.}
\usepackage[margin=1.25in]{geometry}

\newtheorem{hypothesis}{Hypothesis}

\renewcommand{\d}{\mathrm{d}}
\newcommand{\C}{\mathbb{C}}
\newcommand{\N}{\mathbb{N}}
\newcommand{\R}{\mathbb{R}}
\newcommand{\Rone}[1]{#1}%\textcolor{blue}{#1}}
\newcommand{\Rtwo}[1]{#1}%\textcolor{green}{#1}}
\newcommand{\V}{\mathcal{V}}
\newcommand{\tro}{\mathcal{L}}
\renewcommand{\j}{^{(j)}}
\DeclareMathOperator{\sign}{sign}

\newcommand{\htg}{\mathrm{ht}}

\author{Caroline L. Wormell\thanks{Laboratoire de Probabilit\'es, Statistique et Mod\'elisation (LPSM),
		Sorbonne Universit\'e, Universit\'e de Paris \\
		email: {\sf wormell@lpsm.paris}\\
		ORCID: 0000-0003-2953-6493}
	}
\title{\Rone{Non-hyperbolicity at large scales} of a high-dimensional chaotic system}
\begin{document}
	\maketitle

	\begin{abstract}
		The dynamics of many important high-dimensional dynamical systems are both chaotic and complex, meaning that strong reducing hypotheses are required to understand the dynamics. The highly influential chaotic hypothesis of Gallavotti and Cohen states that the large-scale dynamics of high-dimensional systems are effectively uniformly hyperbolic, which implies many felicitous statistical properties. \Rone{We obtain direct and reliable numerical evidence, contrary to the chaotic hypothesis, of} the existence of non-hyperbolic large-scale dynamical structures in a mean-field coupled system. To do this we reduce the system to its thermodynamic limit, which we approximate numerically with a Chebyshev basis transfer operator discretisation. This enables us to obtain a high precision estimate of a homoclinic tangency, implying a failure of \Rone{uniform} hyperbolicity. Robust non-hyperbolic behaviour is expected under perturbation. As a result, the chaotic hypothesis should not be {\it a priori} assumed to hold in all systems, and a better understanding of the domain of its validity is required.
	\end{abstract}

\section{Introduction}\label{s:Introduction}

%GALAVOTTI-COHEN
Most complex systems have chaotic dynamics on a large set of parameters: such systems include those in statistical mechanics, and the Earth's climate system. The chaotic dynamics of such systems being almost universally too complicated to treat from rigorous first principles, general simplifying principles are necessary to understand the system's most important components. When the system is spatially structured, such as in climate models, these components are often the dynamics taking place on large spatial scales \cite{VanVliet08}.

The paradigmatic subclass of chaotic systems are uniformly hyperbolic systems, which have a uniform splitting between expanding and contracting directions \cite{Bowen08}. Because of their simple geometry, these systems are very amenable to study. Without good hyperbolicity assumptions, however, our rigorous knowledge of multidimensional chaotic systems is meagre \cite{Blumenthal17}. This is a major problem, because real-life examples of hyperbolic chaotic dynamics are very rare \cite{Kuznetsov21}, \Rtwo{whereas strong violations of hyperbolicity are common \cite{Lucarini20}}.

Nonetheless, it is conjectured that, when considered at large scales, typical chaotic dynamics resolve as hyperbolic \cite{Gallavotti20}:
\begin{hypothesis}[Gallavotti--Cohen \cite{GallavottiCohen95a, GallavottiCohen95b}] \label{hypothesis}
	The macroscopic dynamics of a (high-dimensional) chaotic system on its attractor can be regarded as a transitive hyperbolic (``Anosov'') evolution.
\end{hypothesis}
In fact, many ``nice'' statistical properties possessed by uniformly hyperbolic systems are also found in the macroscopic-scale dynamics of certain large \Rone{non-uniformly hyperbolic} systems. These properties include existence of physical invariant measures, exponential mixing and large deviation laws \cite{Lebowitz99, Lepri98}. Hypothesised mechanisms include emergent stochastic effects in coupled systems \cite{WormellGottwald18,WormellGottwald19}, matching of topological equivalency classes between different subsystems under perturbations \cite{WormellGottwald19}, and generic distribution of singularities in the system \cite{Ruelle18}. The chaotic hypothesis suggests that general high-dimensional chaotic systems may be studied using techniques developed for uniformly hyperbolic systems, an idea which has been much taken up in the geophysics literature \cite{Gritsun13, Lucarini14}.

However, potential counter-examples arise when considering the response of the physical invariant measure to dynamical perturbations. Hyperbolic systems are known to have a so-called ``linear response'', that is to say their statistics vary differentiably when a parameter of the chaotic system is varied \cite{Ruelle97}: many smooth non-uniformly hyperbolic systems on the other hand fail to have a linear, or even a continuous response \cite{Baladi14, Baladi15}. \Rtwo{This can be traced to the failure of the physical invariant measure's derivative to exist in a function space where the transfer operator decays summably under iteration (e.g. has a spectral gap) \cite{Baladi14, Gouezel06}.} While in many geophysical systems linear response theory has been successful \cite{Bell80, RagoneEtAl15, Lembo19}, certain ones appear to respond non-differentiably to perturbations in broad regimes \cite{Cooper13,Chekroun14}, \Rtwo{where linear response would not be excluded through slow mixing of general smooth functions \cite{Tantet18}}. Nonetheless, \Rone{although it is commonly accepted that the chaotic hypothesis implies some reasonable expectation of structural stability}, it is debatable that linear response falls outside the scope of the chaotic hypothesis, because the hypothesis (\ref{hypothesis}) only pertains to individual systems, whereas a linear response is a property of a family of systems \cite{WormellGottwald18}, \Rtwo{notwithstanding that the existence of a formal linear response candidate is a good indicator of the existence of linear response} \cite{Baladi14}.

%OUR SYSTEM / MEAN FIELD SYSTEMS
%In previous work the chaotic hypothesis applied to linear response was investigated, using the example of globally coupled systems \cite{WormellGottwald19}. 
In recent times, linear response behaviours of complex chaotic systems have been investigated through the increasingly popular model of mean-field coupled maps. These are systems composed of many chaotic subsystems that interact with each other through a mean-field \cite{Kaneko89}. They are a subset of globally-coupled maps \cite{Chazottes05}. As the number of subsystems tends to infinity, the large-scale behaviour of these systems can be described by a so-called thermodynamic limit system \cite{Pikovsky94, Keller00}. These limit systems may exhibit non-trivial and sometimes complex dynamics \cite{Selley19, Ciszak21}. With certain smooth hyperbolic subsystems and sufficiently weak couplings, linear responses have been proven to exist in thermodynamic limit systems \cite{Selley21, Galatolo21}.
On the other hand, \cite{WormellGottwald19} presented a mean-field coupled system whose thermodynamic limit's response to perturbations appeared to be non-smooth. This was argued to be the result of an apparent structural similarity between the thermodynamic limit and the non-uniformly hyperbolic H\'enon map, for which linear response fails.

The goal of this paper is to furnish an explicit example of \Rone{non-hyperbolic structures} in a thermodynamic limit system similar to that of \cite{WormellGottwald19}. The \Rone{non-uniformly hyperbolic} limit system we present has a homoclinic orbit whose stable and unstable directions are tangent to each other, a non-hyperbolic structure which is definitionally excluded in uniformly hyperbolic dynamics \cite{Bowen08}. Because the thermodynamic limit system's attractor contains this homoclinic tangency, it violates Hypothesis~\ref{hypothesis}. 

\Rone{We note that our result does not exclude that, despite the non-hyperbolic structure, the system may still have some kind of non-uniform hyperbolic dynamics on some attractor. However, we only expect hyperbolicity in the dynamics in the sense that Lyapunov exponents exist and are bounded away from zero almost everywhere \cite{Young95}. This is a notably less powerful property than that of Hypothesis~\ref{hypothesis}, and in particular does not imply most of the favourable statistical properties mentioned above.} 
%\red{Might be worth mentioning Ibanez homoclinic tangencies in Navier Stokes}

%HOMOCLINIC TANGENCIES
Apart from self-evidently demonstrating a failure of uniform hyperbolicity for a single limit system, the existence of a generic homoclinic tangency also suggests so-called wild dynamical phenomena on a generic set of nearby limit systems. Examples include the existence of infinitely many sinks ({\it i.e.}~stable periodic orbits), \Rtwo{each with their own basin of attraction} \cite{Newhouse74,Berger16generic,Berger16dimension}, or more relevantly, of other tangencies between stable or unstable manifolds, which is to say the persistence of non-hyperbolic behaviour \cite{Palis94}. %For example,  may also be expected in nearby thermodynamic limit systems, including homoclinic tangencies, and systems possessing infinitely many sinks. 
Thus, the homoclinic tangency we obtained for one system would imply a violation of the chaotic hypothesis for an $O(1)$-size open set of nearby systems.

%NUMERICAL METHODS
Our evidence for the homoclinic tangency is numerical. To approximate the infinite-dimensional limit system, we will apply Chebyshev Galerkin discretisations for the transfer operators that describe the thermodynamic limit \cite{Wormell19, Bandtlow20}, using the software package {\tt Poltergeist.jl} \cite{Poltergeist}. Such discretisations are very accurate and efficient in approximating such objects. To obtain the homoclinic tangency in this system we will use a shooting method. With these methods the homoclinic tangency is estimated to a very high accuracy. This furnishes very strong evidence of its existence.

% high precision, we compute a parameter of the systems containing a generic homoclinic tangency, and demonstrate numerically that the homoclinic is contained in an attractor with chaotic dynamics.We do this using , Poltergeist, etc. These have been found to have very high accuracy (cite Wormell 19)

%{\it Q: would it be worth trying to find a (non-homoclinic) tangency for say $t = 31$? Hard, requires proving existence of blender.}

%RESULTS
%We fin

%CONSEQUENCES
%This poses a clear counter-example to Hypothesis~\ref{hypothesis} in the sense that the macroscopic dynamics are, in fact, non-hyperbolic. %in as much as high-dimensional dynamics can be well-approximated by a low-dimensional system, this system is non-hyperbolic (?). 
%It is theorised that the existence of LRT/structural stability in higher-dimensional systems is not guaranteed, but rather depends on the structure of the system, in various ways that have been or are being elaborated. Outside of the question of perturbations, many low-dimensional non-hyperbolic systems with sufficiently fast decay of correlations are known or believed to possess most important statistical properties, {\it e.g.}~central limit theorems and so on.

The paper is structured as follows. In Section~\ref{s:Model}, the mean-field coupled system and its thermodynamic limit are presented, and in Section~\ref{s:Manifolds} the mathematical objects required to describe a homoclinic tangency are introduced. In Sections \ref{s:Spectral}-\ref{s:Continuation} numerical methods are presented: first the scheme to approximate the thermodynamic limit map, and then the method to find the homoclinic tangency. The results are presented in Section~\ref{s:Results} and in Section~\ref{s:Conclusion} implications and further directions are discussed.

\section{Model}\label{s:Model}

\subsection{Mean-field system}

We introduce a mean-field system, similar to those proposed in \cite{WormellGottwald19}, whose dynamic variables are $M \gg 1$ one-dimensional chaotic subsystems $q\j$ coupled together via a mean field $\Phi$. These chaotic subsystems $q\j \in [-1,1]$ each evolve according to smooth, individually hyperbolic (in fact uniformly-expanding) chaotic dynamics 
\begin{equation} q\j_{n+1} = f_{t\Phi_n}(q\j_n),\label{eq:fdynamics}\end{equation}
modulated by a mean field of the $q\j_n$
\begin{equation} \Phi_n = \frac{1}{M} \sum_{j=1}^M q\j_n \label{eq:Phidef}\end{equation}
and a fixed parameter $t \geq 0$ which determines the strength of the coupling.

%\it{For Pierre:
%\nexttime{
%To rephrase: we consider in the first instance, for $M \in \mathbb{N}$ and $t \geq 0$, high-dimensional maps $\mathbb{F}_{t,M}: [-1,1]^M \circlearrowleft$, with
%\[ \mathbb{F}_{t,M}((q\j)_{j=1\ldots M}) = \left( f_{\frac{t}{M}\sum_{m=1}^M q^{(m)}}(q\j) \right)_{j=1,\ldots,M}\]
%with $\{f_\alpha\}_{\alpha \in \R}$ a family of chaotic maps to be given.\\
%}

\begin{figure}[t]
	\centering
	\includegraphics{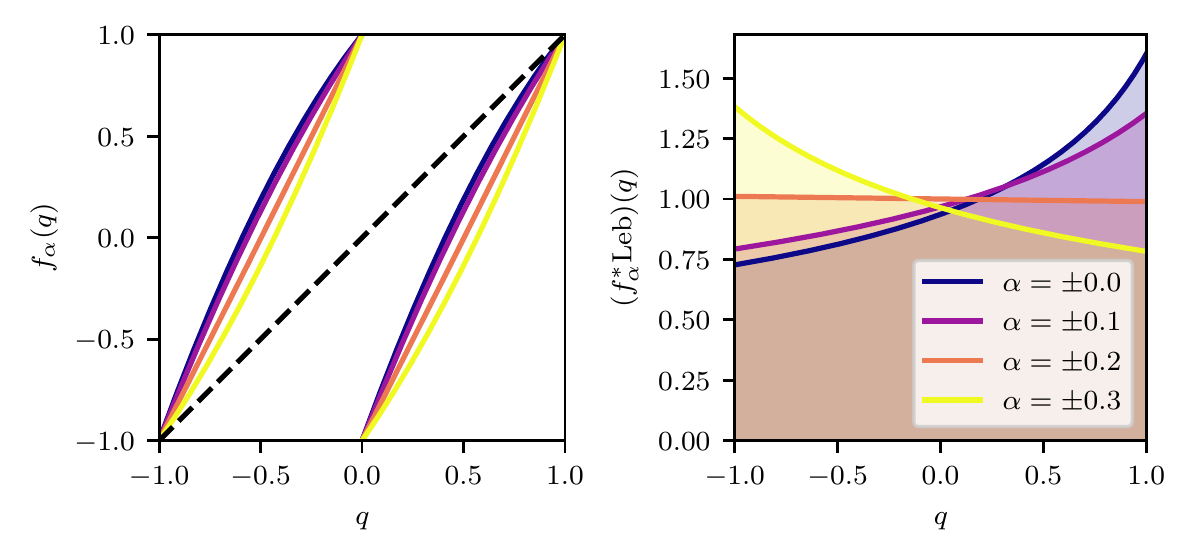}
	\caption[Graph of maps $f_\alpha$]{Top: graph of the microscopic maps $f_\alpha$ for some representative values of $\alpha$. 
		
		Bottom: for various $\alpha$, the action of $f_\alpha$ on Lebesgue measure: that is, $\tro_\alpha 1$ for various values of $\alpha$. Note that because the Lebesgue measure of the domain $[-1,1]$ is $2$, these measures are actually {\it twice} a probability measure.	}
	\label{f:self-coupling-plot}
\end{figure}

For the subsystem dynamics we choose
\begin{equation}\label{eq:fdef}
	f_\alpha(q) = \mathfrak{d}(q) + g(\alpha)(1 - \mathfrak{d}(q)^2),
\end{equation}
where $\mathfrak{d}(q) := 2q - \sign q$ is the doubling map on $[-1,1]$, and
\begin{equation*}
	g(\alpha) = \tfrac{3}{16} \cos 8\alpha \in (-\tfrac{3}{16},\tfrac{3}{16}).
\end{equation*}
These maps, given in Figure~\ref{f:self-coupling-plot}, are piecewise analytic maps of the interval, with two full branches. They are uniformly expanding, with each $|f_\alpha'| \geq \tfrac{5}{4} > 1$.
%The inverses of these family of maps are given by a simple expression
%\[f^{-1}_\alpha(x) = \mathfrak{d}^{-1}\left(x - K_\alpha (1 - \sqrt{0.97 x^2 + 0.03})  \right). \] 
%This speeds computation of the action of the transfer operator, defined later in \eqref{eq:TransferDef}.

Dynamically, the choice of $f$ (in particular of $g$) encourages the $q\j$ to take higher values (increasing $\Phi_{n+1}$) for $\Phi_n$ close to zero, and towards $q = -1$ for $\Phi_n$ appropriately far from zero. 
%This observable, plotted at bottom in Figure~\ref{f:self-coupling-plot}, gives a measure of the centrality of the $q\j$ ensemble: $\phi(q)$ is negative when $q$ is close to zero, and positive when $q$ is near $\pm 1$. 
For $t \lesssim 3$, this induces quasi-unimodal dynamics in the mean position of the $q\j$'s.

\subsection{Thermodynamic limit reduction}

%To reduce this system to its macroscopic components, 
%\nexttime{To reduce the $\mathbb{F}$ system to its macroscopic components,} 
Because the $q\j$ are exchangeable, the macroscopic aspects of the system should be expressible as functions of the distribution of the $q\j$ at fixed times.

%To reduce our system to its macroscopic components (i.e. moments of the distribution of the $q\j$), it
It is in fact possible to use this exchangeability to form an exact closure of the system (\ref{eq:fdynamics}-\ref{eq:Phidef}) in terms of precisely this empirical distribution $\mu_n = \tfrac{1}{M} \sum_{j=1}^M \delta_{q\j_n}$:
\begin{align*} 
	\mu_{n+1} &= (f_{t\Phi_n})_* \mu_n,\\
	\Phi_n &= \int_{-1}^{1} q\, \d\mu_n(q),
\end{align*}
where $(f_{t\Phi_n})_*$ is the push-forward of $f_{t\Phi_n}$. Statistics of the $q\j_n$ can then be recovered through averages over $\mu_n$. As the number of subsystems $M \to \infty$ we can expect the discrete empirical measures $\mu_n$ to converge to probability distributions with appropriately smooth Lebesgue densities \cite{Keller00}. In a mild abuse of notation, we will henceforth use $\mu_n$ to refer to these density functions. We can thus rewrite our dynamics as
\begin{align} 
	\mu_{n+1} &= \tro_{t\Phi_n} \mu_n, \label{eq:ThermoDef1}\\
	\Phi_n &= \varphi\mu_n \label{eq:ThermoDef2}%\int_{-1}^{1} \phi\,\mu_n\, \d q
\end{align}
where $\tro_\alpha$ is the transfer operator of $f_\alpha$, with explicit expression
\begin{equation} (\tro_\alpha h)(x) = \sum_{x \in f_\alpha^{-1}(y)} \frac{h(y)}{f_\alpha'(y)}, \label{eq:TransferDef}\end{equation}
and the functional $\varphi$ is given $\varphi\mu := \int q \mu(q) \, \d q$.

While the system (\ref{eq:ThermoDef1}-\ref{eq:ThermoDef2}) can be reformulated as a delay equation in mean field $\Phi_n$ \Rtwo{using the theory of transfer operator cocycles} \cite{WormellGottwald19}, we will solve it as a function of the measure distribution
\begin{equation} \mu_{n+1} = F_t(\mu_n) := \tro_{t\varphi\mu_n} \mu_n. \label{e:Fsystem}\end{equation}
These maps $F_t$ are our thermodynamic limit systems: they sometimes known as self-consistent transfer operators \cite{Selley19,Selley21}.

Because we have assumed that $\mu_n$ are smooth, absolutely continuous probability measures, we assume that the dynamics takes place within the space $U$ of positive, twice-differentiable densities that integrate to 1 on $[0,1]$. However, we can use the uniform analyticity of the maps $f_\alpha$ to further restrict the thermodynamic limit dynamics $F_t$ to act on a scale of function spaces on which it has relatively nice compactness properties.

\subsection{Hardy function spaces}\label{ss:HardySpaces}

The Banach spaces we use are Hardy spaces $%\{
H_\rho,\, \rho > 0$ of analytic functions. For small enough $\rho$, the sets $H_\rho \cap U$ are invariant sets of the $F$-dynamics on $U$. In fact, for any sufficiently large $C > 0, N \in \N$, the sets $\{h : \|h\|_{H_\rho} \leq C \} \cap U$ are attracting invariant sets\footnote{\Rone{We have that $\mu_n = F_t^n(\mu_0) = \tro_{\alpha_{n-1}}\cdots \tro_{\alpha_0}\mu_n$ for some $\alpha_k$ depending on $\mu_0$. Because $\{\tro_{\alpha_k}\}$ forms a cocycle of transfer operators uniformly bounded in both $H_\rho$ and $C^2$, there exists some $N$ such that each $\tro_{\alpha_{N+n}}\cdots \tro_{\alpha_n}$ is a contraction on both $H_\rho \cap U$ and $C^2 \cap U$ in their respective norms. In particular, $\mu_n$ will converge exponentially in $C^2$ to the time-dependent absolutely continuous invariant measure of the cocycle, which lies in $H_\rho$ and has uniformly bounded $H_\rho$ norm.}} for some iterate $F^N$.

\begin{figure}
	\centering
	\includegraphics{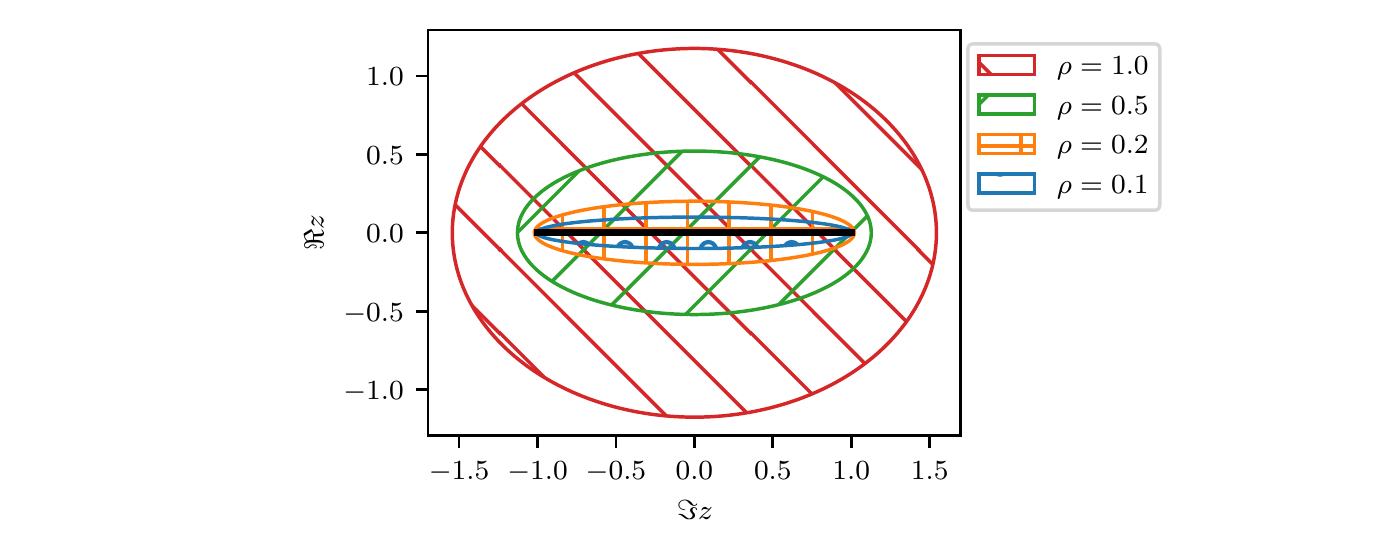}
	\caption{Bernstein ellipses for various parameters $\rho$.}
	\label{f:bernstein}
\end{figure}

The domain of a function in $H_\rho$ is the Bernstein ellipse $E_\rho$:
\[ [-1,1] \subseteq E_\rho = \cos\left([0,\pi] + i[-\rho,\rho]\right) \subset \C. \]
These are ellipses in the complex plane centred at $0$ with semi-axes $\cosh \rho > 1$ and $\sinh \rho > 0$: a range are plotted in Figure~\ref{f:bernstein}. The Hardy space $H_\rho$ is composed of continuous complex functions on $E_\rho$ which are analytic on its interior, equipped with the supremum norm
\[ \| h \|_\rho = \sup_{z \in E_\rho} |h(z)|. \]
As an edge case, the space $H_0$ is simply the space of continuous functions $C^0([-1,1],\mathbb{C})$.

Now, because of its definition through function composition \eqref{eq:TransferDef}, the transfer operator $\tro_\alpha$ is bounded from smaller-$\rho$ Hardy spaces into larger-$\rho$ ones \cite{Bandtlow20}.

In particular, if $0 < R \leq R_{\rm max} := 0.5$ and $r \geq 0.93R$, one can show that all $\alpha \in \R$ that $f^{-1}_\alpha(E_R)$ is a subset of $E_r$. Hence, if $h \in H_r$,
\begin{align}
	\| \tro_\alpha h \|_{R} &= \sup_{z \in E_R} \left| \sum_{w \in f_\alpha^{-1}(z)} \frac{h(w)}{f_\alpha'(w)} \right| \notag\\
	&\leq 2 \sup_{w \in f_\alpha^{-1}(E_R)} \left|\frac{h(w)}{f_\alpha'(w)} \right| \notag\\
	&\leq \frac{2 \sup_{w \in E_r} |h(w)|}{\inf_{w \in f_{\alpha^{-1}(E_R)}}|f_\alpha'(w)|}  \notag\\
	&\leq 1.85 \| h\|_r, \label{eq:HardySpaceTransferNorm}
\end{align}
In particular, every $\tro_\alpha$ is uniformly bounded as an operator $H_r \to H_R$ for $R \in (0,R_{\rm max}]$ and $r \geq 0.93R_{\rm max}$.

This means $\tro_\alpha$ is bounded as an endomorphism on both $H_r$ and $H_R$ since $\| \cdot \|_r \leq \| \cdot \|_R$ for $r \leq R$. Better than this, inclusions between Hardy spaces of different parameters are very strongly compact, in fact nuclear. This is a simple fact falling out of Fourier analysis \cite{Trefethen13}. Hence as an endomorphism on $H_r$, the transfer operator $\tro_\alpha$ has the same strong compactness properties \Rone{uniformly in $\alpha$}.

On these Hardy spaces, the transfer operator also has a very nice perturbation theory. It is a standard fact of complex analysis the differentiation operator $\partial_q$ and its iterates $\partial^k_q$ are also bounded as operators $H_{R} \to H_{r}$ for any $r < R$. A particular consequence of this is that, since the derivatives of the transfer operator with respect to dynamical perturbations, that is $\frac{\d^k \tro_\alpha}{\d \alpha^k}$, can be expressed as linear combinations of $\partial_q^j \tro_\alpha, j \leq k$ \cite[Section~2.3]{Sedro18}, these derivatives are also bounded as operators $H_r \to H_r$. In our setting, the map $\alpha \mapsto \tro_\alpha$ is therefore a $C^\infty$ function $\R \to L(H_r,H_r)$ for appropriate positive choices of $r < 0.93R_{\rm max}$. This justifies the perturbation theory used in the rest of the paper.

%The transfer operators $\tro_\alpha$ are well-defined as endomorphisms on a range of Banach spaces, including spaces of continuous differentiability $C^k$ and Hardy spaces of analytic functions \cite{Baladi08,Bandtlow08}. Because these systems have uniformly bounded uniform expansion and distortion bounds, many relevant quantities of these maps and their cocycles, including the norm, spectral gap and the gap between Lyapunov exponents of the cocycle are uniformly bounded \cite{Baladi96,Bandtlow08, Korepanov15}. Furthermore, they have 

\section{Manifolds and tangencies}\label{s:Manifolds}

It is now necessary to define the manifold structure of $F_t$, which will allow us to speak to its hyperbolicity or absence thereof.

Let us suppose that we have a \Rone{differentiable} dynamical system $F$ acting on an affine subspace $M$ of a Banach space with tangent space $TM$ \Rone{which we may naturally identify with $M \times (M - m^*)$ for any $m^* \in M$}. % with $x \in M$. 
The stable manifold of a point $x \in M$ is the set of points near $x$ whose forward orbits converge to that of $x$:
\[ \V^s_x = \{ y \in M : \lim_{n\to\infty} d_M(F^n x, F^n y) = 0 \}, \]
where $d_M$ is the metric on $M$. The local stable manifold of $x$ is the set of such points which additionally do not leave some small $\delta$-neighbourhood of $x$:
\[ \V^{s,\mathrm{loc}}_x = \{ y \in M : \lim_{n\to\infty} d_M(F^n x, F^n y) = 0, \sup_{n \in \N} d_M(F^n x, F^n y) \leq \delta \} \subset \V^s_x. \]

Similarly, when $F$ is a diffeomorphism, the unstable manifold of $x$ is the set of points with {\it backward} orbits converging to that of $x$:
\[ \V^u_x = \{ y \in M : \lim_{n\to\infty} d_M(F^{-n} x, F^{-n} y) = 0 \}. \]
\Rone{Along hyperbolic trajectories of $x$ and under reasonable conditions, these stable and unstable manifolds are indeed manifolds. Furthermore, if the range of $D_xF$ is dense in $T_x M$ for all $x$, then the global stable manifold $\V^s_x$ has the same codimension as the local stable manifold \cite{Henry06} (we prove this for our maps $F_t$ in the Appendix). On the other hand, if the kernel of $D_xF$ avoids the unstable space then the global unstable manifold $V^u_x$ can be expected to have the same dimension as the local unstable manifold.}

We can extend these notions of stable and unstable manifolds onto the tangent bundles. For $x \in M$ let $D_xF: T M \to TM$ be the differential of $F$, that is to say that for all tangent vectors $v \in T M$
\[ F(x + \epsilon v) = F(x) + \epsilon D_x F v + \mathcal{O}(\epsilon^2). \]
The {\it stable subspace} (resp. {\it unstable subspace}) at $x$, $E^s_x$ (resp. $E^u(x)$) $\subseteq TM$, is then the set of tangent vectors at $x$ which converge to zero under the action of $DF$ (resp. $F^{-1}$):
\begin{align*} E^s_x &= \{ v \in T_x M : \lim_{n\to \infty} D_xF^n\, v = 0 \} \\
	E^u_x &= \{ v \in T_x M : \lim_{n\to \infty} D_xF^{-n}\, v = 0 \}, \end{align*}
where $D_xF$ is the Jacobian (or differential) of $F$.
These are respectively tangent to local stable and unstable manifolds \cite{Bowen08}.

Because our limiting dynamics $F_t$ given in \eqref{e:Fsystem} are not diffeomorphisms,\footnote{Note that, at the expense of complicating the perturbation theory, we could make the maps $F_t$ closer to diffeomorphisms by adding hidden dynamics $r_{n+1} = \tfrac{1}{2}(r_n + \mathbf{1}_{q_n > 1/2}) \in [0,1]$ and choosing $F_t$ to act on an appropriate subset of a Triebel space %$F^{s,t}_p = (I-\Delta_r)^{-t/2}(I-\Delta_q)^{-s/2}(L^p([-1,1]\times[0,1]))$ (defined wrt q,r) for some $1/2>s>-t>0$ 
	\cite{BaladiZetaBook}.} the unstable manifolds and subspaces are ill-defined. However, it is possible to define the unstable manifold (resp. subspace) of a backward orbit $(x_{-n})_{n\in \N}$: this can be achieved using the machinery of natural extensions. If $x_*$ is a fixed point then for convenience we will define $\V^u_{x_*}$ (resp. $E^u_{x_*}$) to be the unstable manifold (resp. subspace) of the orbit $x_{-n} \equiv x_*$.

If $x$ is a fixed point, then $\V^s_x$ are the set of points with orbits converging to $x$, and $\V^u_x$ are the set of points with orbits emanating from $x$; furthermore, provided that the differential $D_xF$ is hyperbolic ({\it i.e.}~its spectrum is bounded away from the unit circle), $E^s_x$ is the span of the stable eigenspaces and $E^u_x$ the span of the unstable eigenspaces.

A separation between unstable and stable subspaces is a key property of most well-behaved chaotic systems. A system is {\it uniformly hyperbolic} if at every point $x \in M$ the tangent space has an $F$-invariant splitting $T_x M = E^s_x \oplus E^u_x$, and there are constants $c > 0$, $\gamma < 1$ such that for all $x \in M$,
\begin{align*}
	\| D_xF^n\, _{| E^s_x}\| &\leq c \gamma^{n},\\
	\| D_xF^{-n}\,_{| E^u_x}\| &\leq c \gamma^{n}.
\end{align*}
According to Hypothesis~\ref{hypothesis}, the $F_t$ are supposedly transitive and (uniformly) hyperbolic on their respective attractors. These two conditions together are the Axiom A of Smale \cite{Bowen08}.

%While Axiom A diffeomorphisms form a fairly restricted subclass of chaotic dynamical systems, % (for example, the only known physical manifestations of these systems are linkage mechanisms), 
%they are widely cited as being representative of the large-scale dynamics of many physical chaotic systems \cite{Gritsun10b, LucariniSarno11, Nijsse19} under the so-called chaotic hypothesis of Gallavotti and Cohen (see Section~\ref{s:lntroduction}).

One generic mechanism to generate non-uniformly hyperbolic dynamics is via homoclinic tangencies. 
%	Non-hyperbolicity is typically very difficult to study, but a simple way to prove its presence in a (family of) systems is to obtain a
%A {\it heteroclinic tangency} in a map $F: M \circlearrowleft$ is an orbit $\{ q_n \}_{n \in \mathbb{Z}}$ of $F$ together with two hyperbolic orbits $\{ p_n \}$ and $\{r_n\}$ such that $p_n - q_n \to 0$ as $n \to -\infty$, and $q_n - r_n \to 0$ as $n \to \infty$ (which is to say that $q_0 \in \V_p^u \cap \V_r^s)$, and furthermore that the stable manifold $ \V_p^u$ and the unstable manifold $\V_q^u$ are tangent. 
A {\it homoclinic tangency} in a map $F: M \circlearrowleft$ is a hyperbolic fixed point $p$ of $F$ together with a different point $q \in \V_p^u \cap \V_p^s$ such that $\V_p^u$ and $\V_p^s$ are tangent at $q$ \cite{Sander00}. In particular, \Rone{because $E_q^{s/u} = T_q V_p^{s/u}$,} the stable and unstable subspaces $E_q^s$ and $E_q^u$ have non-trivial intersection, implying non-hyperbolicity of the given map $F$. 

It is easy enough to show that a homoclinic tangency is equivalent to having that $q \in \V_p^u$ with $\lim_{n \to \infty} F^n(q) = p$ and $\lim_{n \to \infty} \| D_q F^n |_{E_q^u} \| \to 0$, because the stable subspaces $E_\cdot^s$ vary continuously at $p$ since it is a hyperbolic fixed point.

%\red{A {\it heteroclinic tangency} is similar to a homoclinic tangency: instead of a single fixed point one has two (possibly non-fixed) hyperbolic points $p_\alpha, p_\omega$, and one has that $q \in \V_{p_\alpha}^u \cap \V_{p_\omega}^s$ and the two manifolds are unstable. Heteroclinic tangencies also imply non-hyperbolic dynamics.}

\section{Spectral methods}\label{s:Spectral}

In Section~\ref{ss:HardySpaces} we discussed the strong compactness and regularity properties of $F_t$ in certain Hardy spaces. This allows us to very effectively approximate the dynamics of the $F_t$ and invariant manifolds on the computer, through projection onto a basis of Chebyshev polynomials. 

The Chebyshev polynomials $T_k, k = 0, 1, \ldots$ are a polynomial family orthogonal in $L^2_{\rm Cheb} := L^2([-1,1], \d x / \sqrt{1-x^2})$:
\[ \int_{-1}^1 T_k(x) T_j(x) \frac{\d x}{\sqrt{1-x^2}} = w_k \delta_{jk}, \]
where $w_k:= \frac{\pi(1+\delta_{0k})}{2}$.
They have explicit expression
\[ T_k(x) = \cos(k\cos^{-1} x), \]
from which falls out a natural connection to Fourier series under the one-to-two transformation $x = \cos\theta$. 

Chebyshev approximation is well-adapted to this compact inclusion. Let $P_n$ be the $L^2_{\rm Cheb}$ orthogonal projection onto the first $n$ Chebyshev polynomials. It is a standard result that if $C_{R-r} := (1 - e^{-(R-r)})^{-1}$ then for $0 \leq r < R$ we have \cite{Trefethen13}
\begin{equation} \| (I - P_n) h \|_{r} \leq C_{R-r} e^{-(R-r) n} \| h \|_R. \label{eq:HardySpaceInclusionNorm}\end{equation}
%Additionally, if $D$ is the differentiation operator, then there exist constants $C'_{R-r}$ such that
%\[ \| D h \|_{r} \leq C'_{R-r} \| h \|_R. \]

%On the one hand, Chebyshev projection is close in norm to the identity when moving from larger ellipses $E_R$ to smaller ones $E_r$; on the other hand, the transfer operators $\tro_\alpha$ are bounded as operators from certain smaller to certain larger ellipses, {\it i.e.}~

On the other hand, from \eqref{eq:HardySpaceTransferNorm} we know that for $r \leq 0.93R$, $R \in (0,R_{\rm max}]$
\[ \sup_{\alpha \in \R} \|\tro_\alpha h\|_r \leq 0.85 \| h\|_{R}.\]
%for sufficiently small smaller ellipses $r < \rho_{\textrm{max}}$ where $\rho_{\textrm{max}}, R(r) >r$ and $C''_{r,R}$ can be chosen independent $a$
From these two equations we therefore know that for such $r, R$ and for all $\alpha \in \R$,
\[ \| (I - P_n) \tro_\alpha \|_{r} \leq \| I - P_n \|_{r \to R} \| \tro_\alpha \|_{R \to r} \leq 0.85 C_{R - r} e^{-(R - r) n}. \]
This implies that the so-called Chebyshev Galerkin approximation $P_n \tro_\alpha$ of the transfer operator $\tro_\alpha$ converges exponentially to the true operator, and thus so do its spectrum and eigenvalues \cite{Bandtlow20}. This can also be extended to more complex functions of $\tro_\alpha$ such as the differential of $F_t$ defined in \eqref{eq:FJacobian} below.

The Galerkin approximations $P_n \tro_\alpha$ are finite-dimensional, and such operators can be easily computed to high accuracy and faithfully represented using the theory of Chebyshev series \cite{Trefethen13}. In particular, if we represent the image of the projection $P_n$ in the Chebyshev basis $\{T_k\}_{k = 0,\ldots, n-1}$, then the finite-rank operators $P_n \tro_\alpha |_{P_n}$ can be represented as $n \times n$ matrices in this basis, with entries
\[ L^{jk}_\alpha = w_k^{-1}  \int_{-1}^1 (\tro-\alpha T_k)(x) T_j(x) \frac{\d x}{\sqrt{1 - x^2}}. \]
These entries can be computed very efficiently by interpolating the action of the operator on a sufficiently large number of Chebyshev \Rone{nodes, which are cosines of evenly-spaced Fourier nodes}. Indeed, we have exponential convergence in the number of interpolating points $N$ \cite{Trefethen13}
\[ L^{jk}_\alpha = w_k^{-1} N^{-1} \sum_{l=0}^{N-1} (T_j \, \tro T_k)\left(\cos \frac{(2l+1)\pi}{2N}\right) + \mathcal{O}(e^{-R_{\rm max} N}). \]
These sums themselves may be computed for all $j < N$ very quickly using the fast Fourier transform. Further details may be found in \cite{Wormell19}.

In practice, we note that when simulating the $F$ dynamics acting on a specific function $h$, it is often more efficient to approximate the action of $F$ directly
\[ F(h) \approx P_n(\tro_\alpha h) \]
rather than constructing a numerical representation of the operator $P_n \tro_\alpha$ and applying it to $h$. \Rtwo{This is because at every step $\alpha$ and therefore $\tro_\alpha$ is different, and there is therefore no time saved by storing a representation of the transfer operator}.

\section{Numerically obtaining the homoclinic}\label{s:Continuation}

Given that we can compute that action and derivatives of the thermodynamic limit map $F_t$ to high accuracy, it then remains to present a method to compute other dynamical objects associated with the thermodynamic limit. In this section, these objects and the methods to compute them are described, starting from the unstable fixed point (Section~\ref{ss:fixedpoint}), through its local manifold approximations (Section~\ref{ss:localmanifold}) to the homoclinic tangency itself (Section~\ref{ss:localmanifold}).

\begin{figure}
	\centering
	\includegraphics{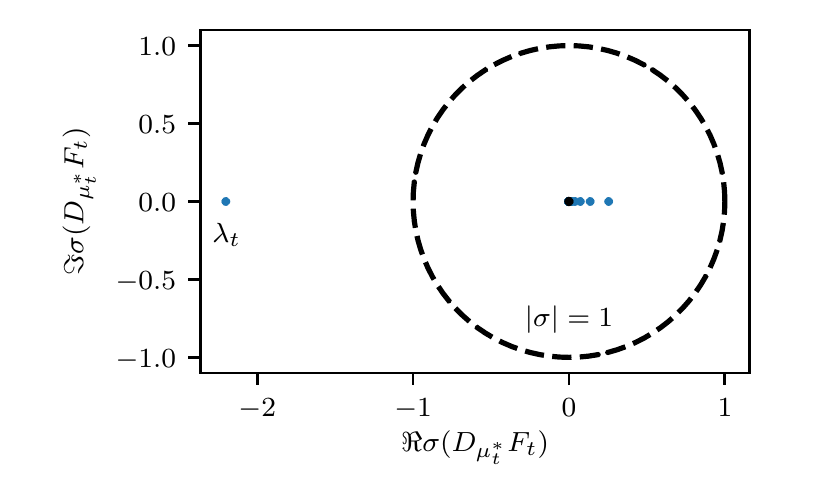}
	\caption{In blue, the \Rone{($\rho$-independent) $H_\rho$ spectrum} of the differential $D_{\mu^*_t}F_t$ of the fixed point of $F_t$, computed for $t = 2.8$ using {\tt Poltergeist.jl}. \Rone{This package uses an adaptive-order Galerkin truncation \cite{Wormell19} and computes eigenvalues via the QR algorithm \cite{ApproxFun}. Away from zero, the spectrum is uniformly stable to perturbations in $t$.}}
	\label{f:jacspec}
\end{figure}

\subsection{Hyperbolic fixed point}\label{ss:fixedpoint}

For \Rone{all $t>0$} the thermodynamic limit system $F_t$ given in \eqref{e:Fsystem} has a \Rone{unique} fixed point $\mu^*_t$ with $\Phi^*_t := \varphi\mu^*_t$ \Rone{lying between $(0,\pi/16t)$}. If we define $\mu_{\rm acim}(\alpha)$ to be the invariant probability density of transfer operator $\tro_\alpha$, then this can be computed numerically\footnote{\Rone{This may be computed via bisection on $\Phi_t^* \in [0, \pi/16t]$. For $t$ close to the homoclinic tangency, a faster-converging method however is to compute $\mu^*_t$ iteratively as a stable fixed point of the relation $\mu_{m+1} = \tfrac{2}{3} F_t(\mu_m) + \tfrac{1}{3} F_t^2(\mu_m)$. Given a sufficiently accurate initial guess, the convergence is justified by considering the Jacobian of this iteration about $\mu^*_t$, which is $D_{\mu^*_t} F_t (\tfrac{2}{3} + \tfrac{1}{3} D_{\mu^*_t} F_t)$. This operator has spectral radius bounded by $0.2$ for $t \approx t_\htg$, which is a result of its spectrum being that of $D_{\mu^*_t} F_t$ under the mapping $z \mapsto \tfrac{z + 2}{3}z$, which almost annihilates the unstable eigenvalue (see Figure \ref{f:jacspec}).}} simply as $\mu^*_t = \mu_{\rm acim}(t\Phi^*_t)$ where $\Phi^*_t$ solves
\[ \Phi^*_t = \varphi\mu_{\rm acim}(t\Phi^*_t). \]
\Rone{This equation also gives the existence and uniqueness of the fixed point, as $\alpha \mapsto \varphi \mu_{\rm acim}(\alpha)$ is strictly decreasing on $[0,\pi/16]$ with a zero at $\alpha = \pi/16$.}

At $\mu^*_t$ the Jacobian of $F_t$ is given by
\begin{equation} D_{\mu^*_t}F_t \psi = \tro_{t\Phi^*_t}\psi - t \varphi \psi \tro_{t\Phi^*_t}\partial_q X_{t\Phi^*_t}, \label{eq:FJacobian} \end{equation}
where $X_\alpha$ is given by
\begin{equation} f_{\alpha+\epsilon}(q) = f_\alpha(q) + \epsilon X_\alpha(f_\alpha(q)) + \mathcal{O}(\epsilon^2) \label{eq:XFunctionalDef}\end{equation}
and is explicitly written in the Appendix. 

Because our $F_t$ dynamics are restricted to probability densities, we restrict the operator $DF_t$ to functions of zero mean in $H_\rho$. Under this restriction, $D_{\mu^*_t}F_t$ is hyperbolic in the sense of that its spectrum \Rone{(which is independent of $\rho$)} uniformly bounded away from the unit circle. In particular, its spectrum outside the unit disc consists of a single eigenvalue $\lambda_t < -1$ (see Figure~\ref{f:jacspec}). This unstable eigenvalue has right eigenfunction $e^u_t \in H_\rho$ and left eigenfunctional $d^u_t \in H_\rho^*$. We normalise $e^u_t$ so that $\|e^u_t\|_{L^2_{\rm Cheb}}=1$ with $\langle 1, e^u_t\rangle_{{\rm Cheb}} > 0$, and we normalise $d^u_t$ to have $d^u_t e^u_t = 1$. We estimate these quantities to near-floating point precision very easily using the Chebyshev spectral methods in Section~\ref{s:Spectral}. In the case of $e^u_t$ this may be done adaptively using {\tt Poltergeist.jl}.

\subsection{Local manifold approximations}\label{ss:localmanifold}

The fixed point's unstable manifold $\V^u_{\mu^*_t}$, which we write for short as $\V^{u,*}_t$, is parametrised near $\mu^*_t$ by 
\begin{equation} \V^{u,*}_t(a) = \mu^*_t + e^u_t a + \tfrac{1}{2} h^u_t a^2 + \mathcal{O}(a^3). \label{eq:VuParametrisation} \end{equation}
where the second-order correction is
\[ h^u_t = ((\lambda_t)^2 - D_{\mu^*_t}F_t)^{-1}H_{\mu^*_t}F_t(e^u_t,e^u_t). \]
The tensor $H_{\mu^*_t}F_t$ is the Hessian of $F_t$ at $\mu^*_t$, with an explicit formula given in Appendix~\ref{appendix}. Like $e^u_t$, the function $h^u_t$ is also easy to accurately approximate with spectral methods.

The parametrisation \eqref{eq:VuParametrisation} can be chosen to have the useful property that $F_t(\V^{u,*}_t(a)) = F_t(\V^{u,*}_t(\lambda_t a))$. Furthermore, the tangent vectors to $\V^{u,*}_t(a)$ are generated by 
\begin{equation} (\V^{u,*}_t)'(a) = e^u_t + h^u_t a + \mathcal{O}(a^2). \label{eq:VuTangent} \end{equation}
At each point $\V^{u,*}_t(a)$ in the unstable manifold, this tangent vector generates its unstable subspace $E^u_{\V^{u,*}_t(a)}$. 

\Rone{It is not {\it a priori} guaranteed that the global unstable manifold has the same dimension as it does near the fixed point\footnote{Although it is generic: the kernel of the differential is smaller than that of the transfer operator (see Appendix), which has infinite codimension.}, but we can produce a dimension-one global manifold from the local manifold by iteration, provided that its tangent vectors avoid the kernel of $DF_t$. In our system, the unstable manifold in fact appears to uniformly avoid the kernel: random sampling of both the unstable manifold and the attractor indicates that there is a bound on the contraction rate 
	\[ \sup_{\mu \in \V^{u,*}_t} \| (DF_t |_{E^u_x})^{-1} \| \leq 0.18^{-1}\]
	for all $t \in [2.7,2.83]$. Numerical investigation suggests the existence of an invariant cone on $TM$ which would yield a robust rigorous bound of this sort on the contraction rate.
}

On the other hand, the fixed point's local stable manifold $p \in \V^{s,\mathrm{loc}}_{\mu^*_t}$, which we write for short as $\V^{s,*}_t$, is close to the kernel of the functional $d^u_t$, so that for $\mu \in \V^{s,*}_t$,
\begin{equation} d^u_t (\mu - \mu^*_t) = \mathcal{O}(\|\mu - \mu^*_t\|_{\Rone{\rho}}^2), \label{eq:Eigenfunctional}\end{equation}
for $\mu \in \V^{s,*}_t$. We expect that, for all local stable points $\mu \in \V^{s,*}_t$ and stable tangent vectors $v \in T_\mu \V^{s,*}_t$, the tangent hyperplanes to such an $\V^{s,*}_t$ satisfy
\begin{equation} d^u_t v = \mathcal{O}(\|\mu\Rone{ - \mu^*_t}\|_{\Rone{\rho}}) \|v\|_{\Rone{\rho}}.\end{equation}
They constitute the stable subspace $E^s_{t,\mu}$ for $\mu \in \V^{s,*}_t$. \Rone{In the Appendix, we show that $D_\mu F_t$ always has dense range on $H_R$, implying that the global stable manifold $\V^s_{\mu^*_t}$ also has codimension one.}

All these quantities can also be accurately estimated via spectral methods. They converge exponentially to the true estimates in the Hardy space $H_R$ for certain $R > 0$, except notably for the leading left eigenfunctional $d^u_t$ of $D_{\mu^*_t}F_t$ at the fixed point, which will converge exponentially in the dual space $H_r^*$ for small $r \in (0,R)$.

We used the Julia package {\tt Poltergeist.jl} to make and adaptively choose the order of the estimates \cite{Wormell19}. The exception again is the left eigenfunctional $d^u_t$, which we computed iteratively via applying $\tro^*$ to a vector of Chebyshev coefficients until convergence was attained. The order of this approximation ({\it i.e.}~the number of Chebyshev coefficients used) was chosen to be approximately that used by {\tt Poltergeist.jl} for estimating the right eigenfunctions\footnote{At higher numerical precisions it was also most efficient here to compute eigendata iteratively via the power method, using {\tt Poltergeist.jl} to compute the action of $D_{\mu^*_t} F_t$ on an eigenvector estimate, renormalising and so on.}.

\subsection{Shooting method}\label{ss:shooting}

Our aim is to find a parameter $t$ and a point $q \in \V^{u,*}_t$ such that $q \in \V^{s,*}_t$ also, with unstable subspace $E^u_{t,q}$ a subset of the stable subspace $E^s_{t,q}$. Because we have good knowledge of the local stable and unstable manifolds near fixed points $\mu^*_t$, we rephrase this as attempting to find a pair $(t,a)$ such that 
\begin{align} F_t^n(\V^{u,*}_t(a)) &\to \mu^*_t, \tag{H1}\label{eq:HomoCond1}\\
	(D_{\V^{u,*}_t(a)}F_t^n)(\V^{u,*}_t)'(a)&\to 0. \tag{T1}\label{eq:TangCond1}\end{align} 
Since $F_t^n(\V^{u,*}_t(a))$ converges towards $\mu^*_t$, we can use the local linearisation of the stable manifold \eqref{eq:Eigenfunctional} and (\ref{eq:HomoCond1}--\ref{eq:TangCond1}) becomes equivalent to solving for $(t,a)$
\begin{align} d^u_t (F_t^n(\V^{u,*}_t(a)) - \mu^*_t) &\to 0, \tag{H2}\label{eq:HomoCond2}\\
	d^u_t (D_{\V^{u,*}_t(a)}F_t^n)(\V^{u,*}_t)'(a)&\to 0, \tag{T2}\label{eq:TangCond2}\end{align} 
where now the left-hand quantities are one-dimensional instead of lying in a Banach space $H_r$ as before.

One way to satisfy these conditions is to find for each $t$ an $a(t)$ satisfying \eqref{eq:TangCond2}, and then finding a $(t,a(t))$ satisfying \eqref{eq:HomoCond2}. \Rone{This turns out to be the most numerically stable option, because when done in this order the roots of both problems are simple and isolated, which is not the case for other approaches\footnote{To lowest order, the left-hand side of \eqref{eq:HomoCond2} is $\sim (a-a_\htg)^2 + (t - t_\htg)$ and the left-hand side of \eqref{eq:TangCond2} is $\sim (a-a_\htg) + (t-t_\htg)$.}.} In both cases, because of the finite relative precision $\varepsilon$ of floating-point arithmetic, and because $\mu^*_t$ is a saddle, we will not in practice be able to compute an orbit converging to $\mu^*_t$ by simple shooting. Instead, we must assume that the convergence holds along the orbit up until some $n = n_*$ determined by the floating-point precision, and \Rone{do some careful error analysis based on the blowup of this error in $\epsilon$.}

We eventually chose our floating-point precision to be 159 bits ({\it i.e.}~three times as many bits as the standard double precision), using the GNU MPFR library implemented as the {\tt BigFloat} type in Julia: \Rone{in particular, the relative precision of the floating-point encoding is $\epsilon \approx 2.7 \times 10^{-48}$}. Having progressively refined our guess at lower precisions, as bracketing intervals we chose 
\begin{align*} a &\in \numprint{0.79276022950246490} + [0,2\times10^{-17}]\\
	t &\in \numprint{2.786033304650978791} + [0,2\times 10^{-18}].\end{align*}

To compute $q_t = \V^{u,*}_t(a)$ accurately we apply the approximation \eqref{eq:VuParametrisation} to $F^{-n_\epsilon}(q_t) = \V^{u,*}_t(\lambda^{-n_{0}}a)$, where we choose $n_{0} = \lceil \log_{2.19}(\epsilon^{-1/3}) \rceil = 16$. Because $\lambda_t \approx -2.19$ for $t \approx 30$, this gives us an error $|F^{-n}(q_t) - \mu^*_t| = \mathcal{O}(\epsilon^{2/3})$. Because we are shrinking our starting point by $\mathcal{O}(\epsilon^{1/3})$, it also gives us an effective numerical precision of $\mathcal{O}(\epsilon^{2/3})$ rather than the full $\mathcal{O}(\epsilon)$.

To estimate the tangent vector at $q$, $(\V^{u,*}_t)'(a)$, we estimate 
\[(\V^{u,*}_t)'(a) = \lambda_t^{-2n_{0}} \left(D_{\V^{u,*}_t(\lambda_t^{-n_{0}} a)}F^{n_{0}}_t\right)\, \left(D_{\V^{u,*}_t(\lambda_t^{-2n_{0}} a)}F^{n_{0}}_{t}\right)\,(\V^{u,*}_t)'(\lambda_t^{-2n_{0}}a)\]
where $\V^{u,*}_t$ and its derivative are computed using \eqref{eq:VuTangent} when they are evaluated at $\lambda_t^{-2n_{0}} a$. This also returns an error of $\mathcal{O}(\epsilon^{2/3})$.

For fixed $n = n_1$, the left-hand sides of \eqref{eq:HomoCond2} and \eqref{eq:TangCond2} are monotone in $a$ and $t$ over sufficiently small intervals. We therefore fixed $n$ in these two equations and used interval subdivision over parameters of $t$ to find $(t,a(t))$ satisfying \eqref{eq:HomoCond2}, where for each $t$ we found (again by subdivision) $a(t)$ satisfying \eqref{eq:TangCond2}. 

We fixed $n_1 = 5 + \lceil \log_{2.19\times 0.255^{-2}} \epsilon^{-2/3} \rceil = 12$, which is approximately when we expect the quantities in \eqref{eq:HomoCond2} and \eqref{eq:TangCond2} to reach their minimum within the bounds of our numerical precision. This choice can be explained as follows. The homoclinic orbit $F^n(\V^{u,*}_t(a))$ approaches $\mu^*_t$ as $\mathcal{O}(\tilde \lambda^n_t)$, where $\tilde \lambda_t \approx 0.255$ is the spectral radius of the fixed point differential $D_{\mu^*_t}F$ in the stable subspace. Because $d^u_t$ captures the local stable manifold of $\mu^*_t$ to first order, it has a second-order error in the distance to the fixed point, and so the quantities in \eqref{eq:HomoCond2} and \eqref{eq:TangCond2} decay as $\mathcal{O}(0.255^{2n})$. On the other hand, the initial error grows as $\mathcal{O}(\epsilon^{2/3} \lambda_t^n)$. This halts the decay of the quantities we are interested in at $n \approx n_1$, and the magnitudes of these quantities bottom out at $\epsilon^{2/3(1+\log 0.255^{-1} / 2\log 2.19)} = 3 \times 10^{-20}$.

Our shooting method (as well as routines to compute stable and unstable manifolds in extended floating point) is contained in the supplementary file {\tt quadratic3.jl}.

\section{Results}\label{s:Results}

We can now present the results we used to compute these quantities.

\begin{figure}
	\centering
	\includegraphics{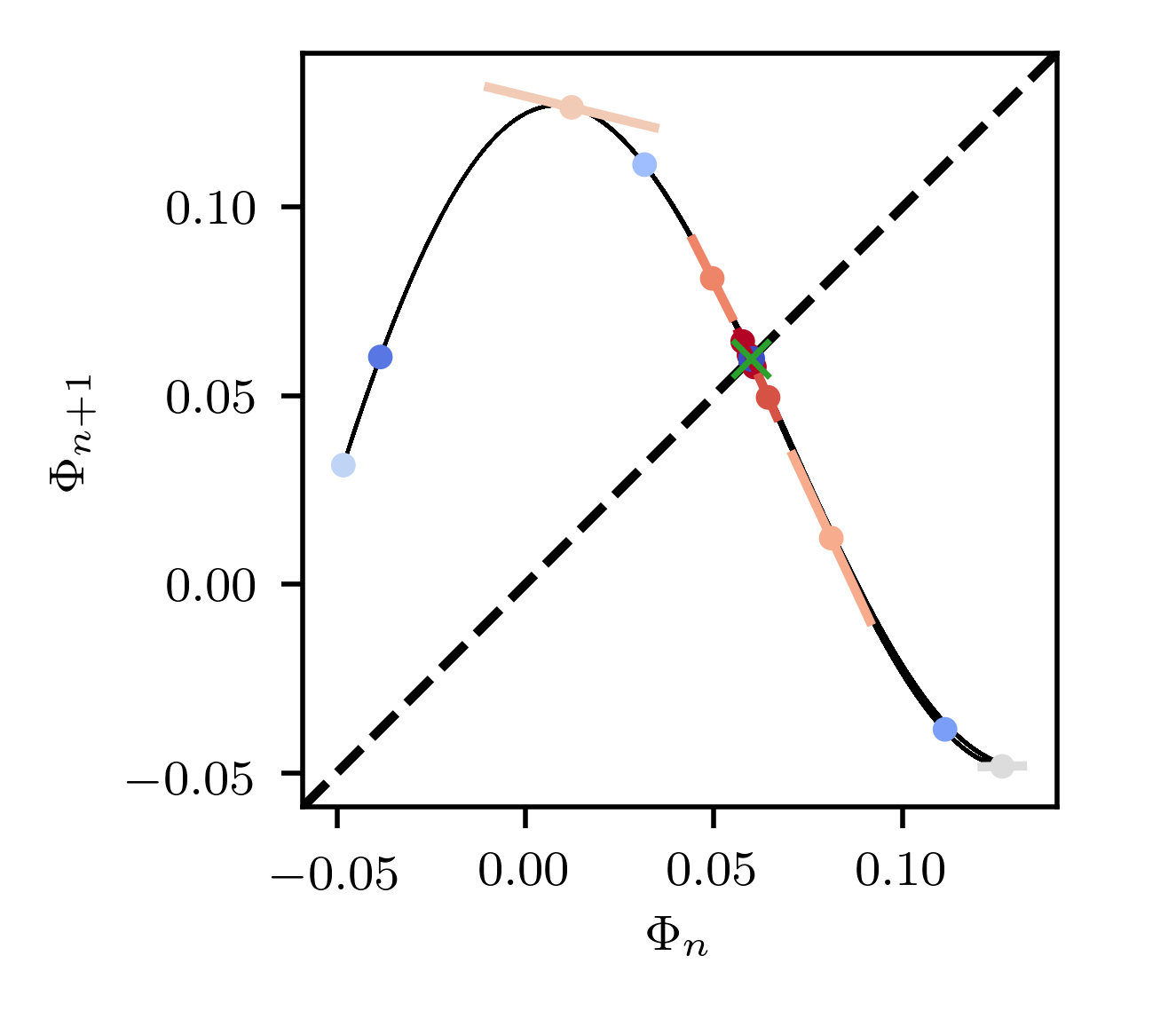}\includegraphics{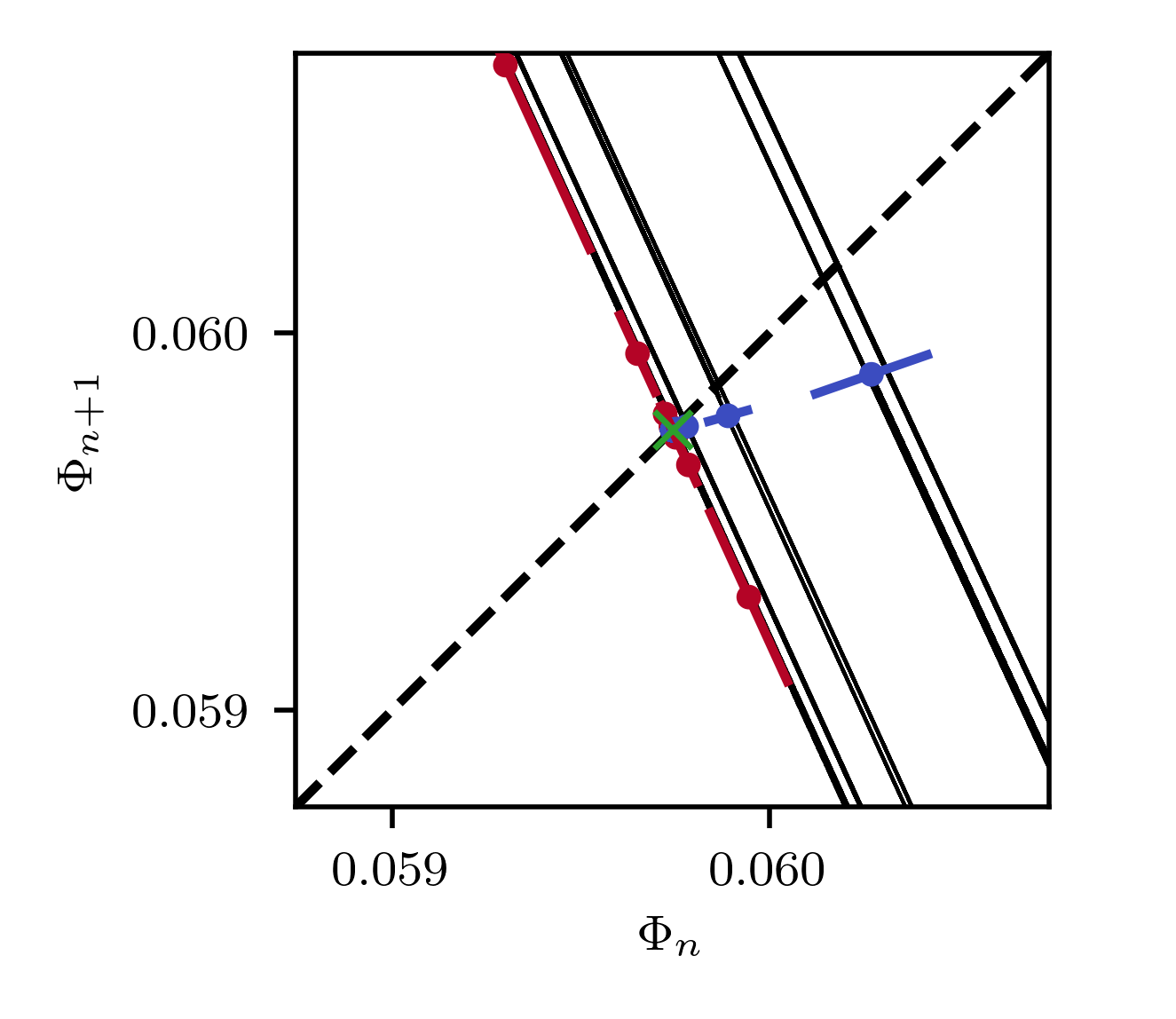}
	\caption[The homoclinic orbit]{Left: \Rone{plot of the homoclinic orbit (dots) and its covariant unstable vectors (lines centred on dots), plotted in colours from red ($n$ small) to blue ($n$ large)}, projected onto delay coordinates in the mean-field $\Phi_n$. The attractor of the $F$ dynamics is plotted in black, and the fixed point $\mu^*_{t_\htg}$ is plotted as a green cross. \Rone{Note that the unstable vectors are as expected all tangent to the attractor.}
		
		Right: detail near the fixed point. The homoclinic orbit \Rone{emanates from (red) and falls into (blue)} the fixed point, with the unstable vector clearly expanded (resp. contracted) by the dynamics.
	}
	\label{f:htorbit}
\end{figure}

\subsection{Existence of a homoclinic tangency}\label{s:existence}

Using the shooting method we obtained the following (non-rigorously validated) estimates for the parameters of a homoclinic tangency:
\begin{align*}
	t_\htg &= \numprint{2.7860333046509787921845397094842} \pm 2 \times 10^{-31} \\
	a_\htg &= \numprint{0.7927602295024649096174830885825} \pm 2 \times 10^{-31}.
\end{align*}

\begin{figure}[h]
	\centering
	\includegraphics{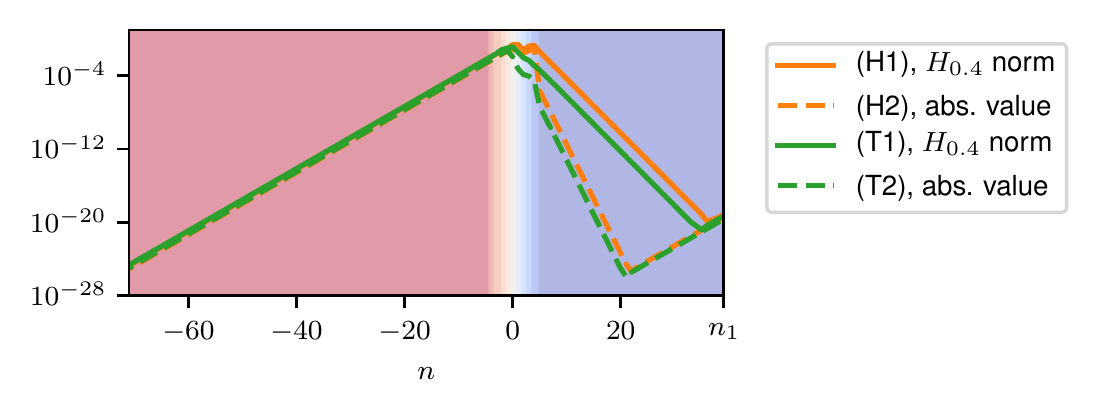}
	%	\caption{The displacement $F^n_t(\mathcal{V}^u_t(a)) - \mu^*_t$ from points along the numerical approximation of the homoclinic orbit to the linearised stable manifold of $\mu^*_{t_\htg}$ is plotted: in Hardy space norm (\eqref{eq:HomoCond2}, orange line) and in unstable projection (\eqref{eq:HomoCond1}, orange dashes). The corresponding covariant unstable vector $d^u_t D_{\mathcal{V}^u_t(a)}F^n_t((\mathcal{V}^u_t)'(a))$ of $\mu^*_{t_\htg}$ is also plotted: in Hardy space norm (\eqref{eq:TangCond2}, green line) and in unstable projection (\eqref{eq:TangCond1}, green dashes). }
	\caption{The norm of the left-hand side of various equations (which theoretically should converge to zero as $n\to\infty$) for our numerical approximation of the homoclinic tangency. Background colours correspond to those in Figure~\ref{f:htorbit}.}
	\label{f:htvalues}
\end{figure}

The relevant homoclinic orbit is plotted with its unstable vectors in Figure~\ref{f:htorbit}, and the quantities (\ref{eq:HomoCond2}--\ref{eq:TangCond2}) we aimed to minimise are plotted in Figure~\ref{f:htvalues}. In Figure~\ref{f:htvalues} we also verify that the original homoclinic tangency conditions (\ref{eq:HomoCond1}--\ref{eq:TangCond1}) are also satisfied.

The precision achieved in these estimates is of a comparable $\epsilon^{2/3} = 2 \times 10^{-32}$ error with our $159$-bit floating point precision. The quantities in \eqref{eq:HomoCond2} and \eqref{eq:TangCond2} reach their minima a little before $n = n_1 = 12$, as predicted, and these minima are of the order of $\epsilon^{2/3(1+\log 0.255^{-1} / 2\log 2.19)} = 8 \times 10^{-21}$, also as predicted. 

These results are obtained to a high precision, with all apparent errors being of the predicted order. Because the thermodynamic limit system $F_t$ has very high regularity with strong compactness properties, it is therefore essentially guaranteed that such a homoclinic tangency exists.
\\

We also have evidence that the homoclinic tangency is generic in two ways that together suggest persistent wild, non-hyperbolic behaviour under perturbation \cite{Berger16generic}.

\begin{figure}
	\centering
	\includegraphics{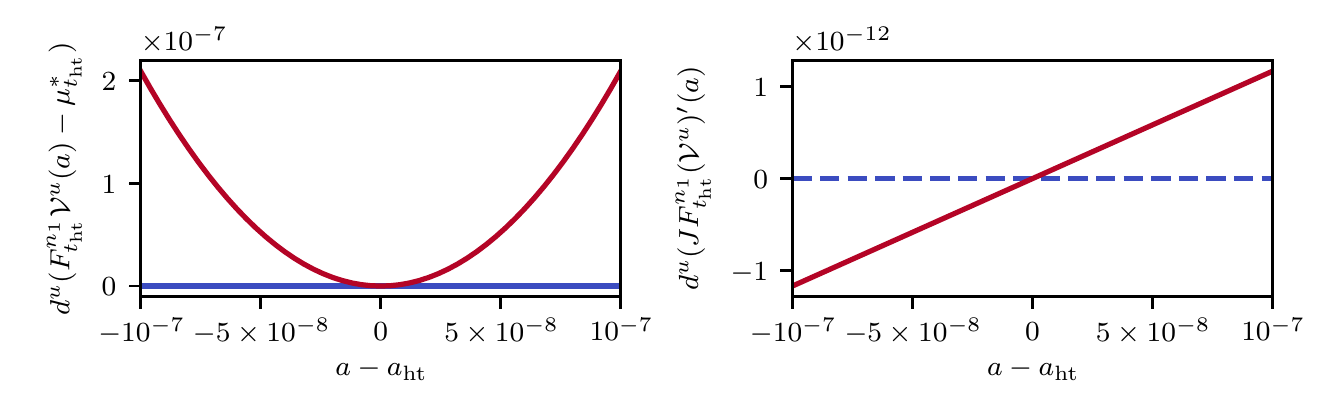}
	\caption[The homoclinic tangency is quadratic]{Left: the unstable manifold near the homoclinic point $F^{n_1}(\V^u_{t_\htg}(a_\htg))$ (red) and the linearised stable manifold (blue). These are projected to one dimension with the eigenfunctional $d^u_{t_\htg}$. 
		
		Right: the same projection of the unstable manifold's derivative $\frac{d}{da}F^{n_1}(\V^u_{t_\htg}(a_\htg))$ (red) and of the linearised stable manifold (blue).}
	\label{f:quadratic}
\end{figure}

Firstly, the homoclinic tangency is quadratic, that is to say that the tangency between the stable and unstable manifolds is a quadratic tangency. The functional $d^u_t$ measures the distance to the local stable manifold of $\mu^*_t$ (to first order in the distance from $\mu^*_t$): in Figure~\ref{f:quadratic} we plot the distances to the unstable manifold at $F^n_{t_\htg}(\V^u_{t_\htg}(a))$ for $a \approx a_{\htg}$, as well as the derivative of this with respect to $a$. The derivative is smooth and clearly has non-zero slope at $a = a_{\htg}$, meaning the tangency is quadratic. This could be explicitly demonstrated in future work by showing $\frac{\d^2 }{\d a^2}F^{n_1}_{t_\htg}(\V^u_{t_\htg}(a)) \neq 0$.

\begin{figure}
	\centering
	\includegraphics{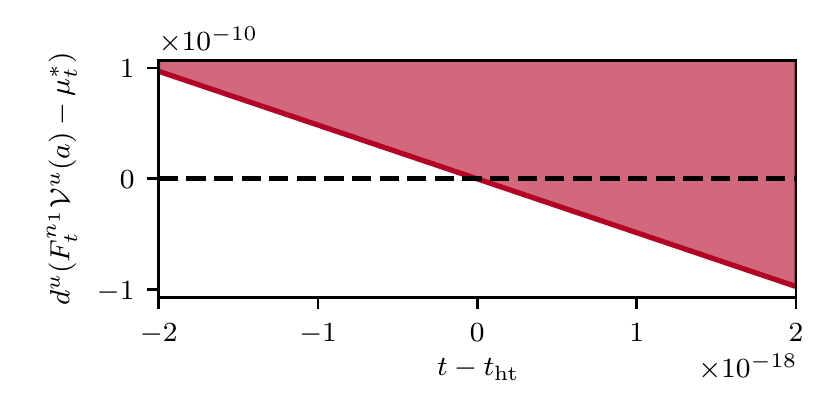}
	\caption{In block colour, the displacement from the local unstable manifold to the local stable manifold at $F^{n_1}(\V^u_{t}(a))$ for $a \approx a_{\htg}$ and $t \lesssim t_{\htg}$, projected onto one dimension using $d^u_t$. As a line, the minimum value attained for given $t$.}
	\label{f:transversality}
\end{figure}

Secondly, we also have strong evidence that tangency is perturbed generically, in the sense that for all $t$ slightly less than $t_\htg$ the unstable manifold around the homoclinic orbit is locally separated from the stable manifold (see Figure~\ref{f:transversality}). In fact, as we might expect, the displacement of the unstable manifold as $t$ is varied is linear and transversal to the stable manifold. If our system was a diffeomorphism, satisfaction of these criteria for a quadratic tangency would imply the existence of heteroclinic tangencies ({\it i.e.}~\Rone{non-uniform hyperbolicity}) on an open set of parameters $t$ \cite{Newhouse79, Palis94}.

To this end, we find that at $t = t_\htg$, the fixed point is {\it sectionally dissipative}: the differential of the fixed point has leading eigenvalue $\lambda_t < - 1.9898$ and all other eigenvalues having modulus less than $0.255 < |-1.9818|^{-1}$, meaning that any product of two eigenvalues has modulus less than one. As a result, we can also expect a Baire~generic Cantor set of parameters $t$ where an infinite number of stable periodic orbits coexist \cite{Palis94, Berger16dimension}

%\red{
%Furthermore by a result of \cite{Berger16generic}, refining \cite{Palis94} would have, in finite dimensions, that the homoclinic tangency will create a Baire generic set of parameter values on which there are an infinite number of stable periodic orbits.
%}

\subsection{Dynamics at $t = t_\htg$}

Although we have that the map $F_t$ is \Rone{non-uniformly hyperbolic}, Hypothesis~\ref{hypothesis} applies only to dynamics on the attractor of the system, which is to say, presumably, on the attractor of the macroscopic dynamics. We therefore wish to have some idea of the macroscopic dynamics's attractor, and in particular whether homoclinic tangency lies on it.

We simulated the thermodynamic limit dynamics at $t=t_\htg$ using {\tt Poltergeist.jl} \cite[Appendix~B]{WormellGottwald19}. The dynamics is chaotic: through simulations on 10 time series of $10^4$ realisations we estimated the leading Lyapunov exponents as $\lambda_1 = 0.381 \pm 0.005 > 0$, $\lambda_2 = -1.150 \pm 0.003 < 0$, $\lambda_3 = -2.015 \pm 0.004 < 0$. \Rone{The existence of a positive Lyapunov exponent suggests the attractor has dimension at least $1$}: the estimates on the Lyapunov exponents then suggest that its Kaplan-Yorke dimension is $1.249 \pm 0.002$. The dynamics also appears to be exponentially mixing over a timescale $t_{1/e} \approx 9$. \Rone{Because there appear to be Lyapunov exponents that are well-defined and away from zero, it is reasonable to claim that the dynamics on the attractor are indeed non-uniformly hyperbolic, if only in a rather weak sense \cite{Young95}.}

\begin{figure}
	\centering
	\includegraphics{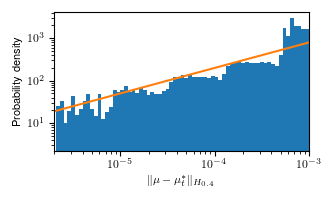}
	\caption[Histogram of distances from the fixed point]{In blue, a histogram of the Hardy space $H_{0.4}$ distance from fixed point $\mu^*_t$ in the attracting $F_t$-dynamics. In orange, a logarithmic slope of gradient $0.6$, indicating a local fractal dimension of around $1.6$ for the SRB measure of $F_t$ at the fixed point. The histogram was obtained from a single time series $\mu_n$ of which $400,000$ timesteps had $d_{H_{0.4}}(\mu_n,\mu_t^*) \leq 0.003$.}
	\label{f:disthist}
\end{figure}

Under this assumption that the attractor is chaotic, there is substantial evidence that the fixed point's unstable manifold, and hence the homoclinic orbit actually lies on this attractor. In Figure~\ref{f:htorbit} it is clear that the attractor contains long unstable manifolds that pass near the fixed point in a direction generally parallel to the unstable vector of the fixed point. We therefore expect that these unstable manifold must have intersections with the stable manifold of the fixed point, implying that the fixed point lies on the attractor.
%Given that, there is substantial numerical evidence to expect the unstable fixed point, and hence the homoclinic orbit, actually lies on this fractal attractor. 
To support this, we took a long time series of $F_t$ dynamics and collected the $H_{0.4}$ distance of points in the time series to the fixed point $\mu^*_t$ (see Figure~\ref{f:disthist}). There is a regular scaling of the physical measure of the $F_t$ dynamics as the distance from the fixed point tends to zero corresponding with a measure dimension at the fixed point of around $1.6$. The closest point in the time series was $6 \times 10^{-7}$ away from the fixed point in $H_{0.4}$ distance, which is of the order expected when the true minimum distance to the attractor is zero. 

Because the attractor contains the fixed point and the system is chaotic (so cannot be confined to the stable manifold of the fixed point), the attractor must also contain the unstable manifold of the fixed point and thus the homoclinic tangency. As a consequence, we can conclude that, not just the map $F$, but the actual {\it large-scale dynamics on the attractor} are non-uniformly hyperbolic. This is in contradiction of Hypothesis~\ref{hypothesis}.
% Because the homoclinic tangency is contained within the attractor, then we know that if the macroscopic dynamics have an SRB measure then they are distributed along the unstable manifold of the system

\section{Conclusion}\label{s:Conclusion}

The chaotic hypothesis (Hypothesis~\ref{hypothesis}) makes a broad claim about the large-scale behaviour of complex chaotic systems. This paper provides a counterexample to it in a mean-field coupled system. We studied this system's thermodynamic limit, which encodes the large-scale dynamics and which the chaotic hypothesis therefore predicts to be hyperbolic on its attractor. We have found however that this attractor contains a homoclinic tangency, and the dynamics on it therefore are \Rone{non-uniformly hyperbolic}. 

On the other hand, a commonly cited restriction on Hypothesis~\ref{hypothesis} is that it holds only for {\it generic} systems: thus, the homoclinic tangency we find could be a special point. However, as a result of having certain genericity properties discussed in Section~\ref{s:existence}, our homoclinic tangency can be expected to generate non-hyperbolic structures on an open set of parameters, following a result of \cite{Palis94}. This results state that for $C^\infty$ diffeomorphisms with one unstable direction but perhaps infinite stable directions, these genericity properties imply the existence of heteroclinic tangencies between stable and unstable manifolds of (non-fixed) hyperbolic points, on an open set of nearby parameters\footnote{\Rtwo{Similar results extending finite-dimensional results with more unstable directions \cite{Bonatti03} may also hold in this infinite-dimensional setting, although they may be difficult to prove.}}  \cite{Palis94}. While our system is not a diffeomorphism, this arises because of very strong stable contracting directions and so fits the spirit of the result. Thus, as well as having a non-uniformly hyperbolic system at one parameter $t = t_\htg$, we in fact expect a failure of uniform hyperbolicity on an interval of parameters \Rone{near $t = t_\htg$}.

Furthermore, homoclinic tangencies can birth a wide array of exotic objects \cite{Bonatti96,Mora93,Yorke83,Palis94}. Of particular note is the Newhouse phenomenon, where infinitely many sinks may coexist in a single system: because the fixed point of the homoclinic tangency is sectionally dissipative, this peculiar phenomenon occurs on a Baire generic\footnote{A Baire generic set is the countable intersection of open dense sets. Baire genericity does not imply a set is of positive measure. For diffeomorphisms of two-dimensional manifolds, the Hausdorff dimension of the set of Newhouse parameters is at least $1/2$ \cite{Berger16dimension} and is conjectured to be of measure zero \cite{Palis08}.} set of parameters $t$ close to $t_\htg$ \cite{Palis94, Newhouse74}. It is a curiosity therefore that for these Newhouse parameters, it is impossible to sample all possible (if unlikely) attracting dynamics. On the other hand, passing from the thermodynamic limit back to finite ensemble size $M$ introduces a small Gaussian noise that will immediately break such dynamical structures.

%While we have very good evidence for a failure of hyperbolicity in our mean-field coupled system, 
Although they demonstrate a kind of generic non-uniform hyperbolicity in globally coupled systems, these coupled systems are somewhat atypical of many real-world systems in the sense that they have a uniform all-to-all network structure. In many more realistic systems such as in geophysics, interactions may be spatially localised, and separation between spatial scales may be incomplete. In such systems, emergent noise that forms the first-order correction to the thermodynamic limit may lead to large-scale {\it stochastic} dynamics \cite{WormellGottwald18, WormellGottwald19}, which share many helpful similarities with hyperbolic dynamics \cite{Young19}. Furthermore, even in a globally-coupled systems, physically meaningful coupling types such as attractive or repulsive behaviours may impose their own dynamical constraints which preclude non-uniformly hyperbolic behaviour \cite{Koiller10}. Nonetheless, our results show that one cannot guarantee ``nice'' large-scale dynamics: these depend on the structure of the complex system.

%Implications: wild dynamics {\it e.g.}~Newhouse phenomenon. Homoclinic tangencies on a dense set. Etc. We caution that these results are for finite-dimensional systems, but the $F$-dynamics on the Hardy spaces are sufficiently contracting on high-order modes that we expect many of the finite-dimensional results to carry over to our setting.

%Are are results really true? Not rigorous but good reasons to believe yes. May be possible to rigorize. Technical questions about can you get a diffeomorphism.

\section*{Acknowledgements}

This research has been supported by the European Research Council (ERC) under the European Union's Horizon 2020 research and innovation programme (grant agreement No 787304).

The author would like to express her thanks to Georg Gottwald for many helpful discussions. Thank you also to Giovanni Gallavotti for his perspective on the chaotic hypothesis and to Roberto Castorrini for comments on the manuscript.% and to .

\appendix
\section*{Appendix}\label{appendix}

\subsection*{$X^\alpha$ and derivative operators}

The perturbation function of $f_\alpha$ as defined in \eqref{eq:XFunctionalDef} is given explicitly by
\[ X_\alpha(q) = g'(\alpha) (1 - \mathfrak{d}(f^{-1}(q))^2) = g'(\alpha)\left(1 - \left(\frac{2(q-g(\alpha))}{1 + \sqrt{1 - 4g(\alpha)(q - g(\alpha))}}\right)^2\right) \]
Let $X^{(1)}_\alpha :=\frac{\partial X_\alpha}{\partial\alpha}$ be its derivative with respect to $\alpha$.

If we define the operator $\Gamma_t(\mu) := \tro_{t\varphi\mu}$ then we have $F_t(\mu) = \Gamma_t(\mu)\mu$. It is standard \cite{Baladi14} that the operator
\[ D_\mu\Gamma_tv = (\varphi v) \partial_q X_{t \varphi\mu} \Gamma_t(\mu) \]
and by recursion, the Hessian
\[ H_\mu\Gamma_t(v,w) = (\varphi v) \partial_q \left( X^{(1)}_{t \varphi\mu} \Gamma_t(\mu) + X_{t \varphi\mu} (D_\mu\Gamma_t v)\right). \]

As a result, the differential and Hessian of $F$ are respectively given by
\begin{align*} (D_\mu F_t)(v) &= \Gamma_t(\mu) v + D_\mu(\Gamma_t v) \mu \\
(H_\mu F_t)(v,w) &= \left(D_\mu\Gamma_t(v + w)\right) \mu + H_\mu\Gamma_t(v,w) \mu.
\end{align*}

Further derivatives of the operator $\Gamma_t(\mu)$ (and hence of $F_t(\mu)$) can be obtained from \cite{Sedro18}.
%\[ \frac{\partial X_\alpha}{\partial \alpha}(x) = g'(\alpha)  \] 

\subsection*{\Rone{Proof that the range of $D_\mu F: H_\rho \circlearrowleft$ is dense}}

\Rone{We first prove that the range of $\tro_\alpha: H_\rho \circlearrowleft$ is dense for any $\alpha \in \R$. Let us first note from \eqref{eq:fdef} that we can decompose $f_\alpha = m_\alpha^{-1} \circ \mathfrak{b}$ where $\mathfrak{b}(x) := x \mod 1$, such that $m_\alpha: E_\rho \mapsto m_\alpha(E_\rho)$ is biholomorphic for some sufficiently small $\rho>0$. For these $\rho>0$, transfer operator $\tro_\alpha$ can therefore be shown to have the form}
\[ \tro_\alpha h = w_\alpha\ ((I + \mathcal{B})h) \circ m_\alpha \]
\Rone{for some weight function $w_\alpha: E_\rho \to \C\backslash\{0\}$, where $(\mathcal{B}\chi)(z) := \chi(z-1)$. We would like to show that for any $\psi \in H_\rho$ and $\epsilon > 0$ there exists $h \in H_\rho$ such that $\| \psi - \tro_\alpha h \|_{H_\rho} \leq \epsilon$. This is to say that we would like to find an $h$ such that the following is small:}
\begin{equation} \sup_{z \in m_\alpha(E_\rho)} \left| \left((I + \mathcal{B})h\right)(z)- \frac{\psi(m_\alpha^{-1}(z))}{w_\alpha(m_\alpha^{-1}(z))} \right| \label{eq:Approximand}\end{equation}

\Rone{By the Stone-Weierstrass theorem, there exist a series of polynomials $\{p_n\}_{n\in \N}$ approximating the continuous second term in \eqref{eq:Approximand} arbitrarily closely on $m_\alpha(E_\rho)$. If $\mathcal{E}_\delta(z) := e^{\delta z}$, it is also possible to choose $\delta >0$ such that $\mathcal{E}_\delta - 1$ is arbitrarily small on $m_\alpha(E_\rho)$. Let us therefore choose}
\[ h = \sum_{i=1}^\infty (-\mathcal{B})^i \mathcal{E}_\delta p_n. \]
\Rone{This function lies in $H_\rho$ because $\mathcal{E}_\delta p_n$ is entire and decays exponentially as $\Re z \to -\infty$, uniformly for bounded $\Im z$. Furthermore, $(I+\mathcal{B}) h = \mathcal{E}_\delta p_n$, giving us what is required, that is, the density of the range of $\tro_\alpha: H_\rho \circlearrowleft$.}

\Rone{We now show that $D_\mu F_t(H_\rho) \supseteq \tro_\alpha(H_\rho)$, which will deliver us the density of the range of $D_\mu F_t$. We have from \eqref{eq:FJacobian} that }
\[ D_{\mu} F_t h =  \tro_{t\varphi \mu}h - \varphi h\ t\tro_{t\varphi \mu}\partial_q X_{t\varphi \mu}. \] 
\Rone{Note now that $u(x) = \tfrac{\pi}{2} \sin \pi x$ lies in $H_\rho$ with $\varphi u = 1$, and, because it is odd, lies in the kernel of all the $\tro_\alpha$. Hence, for any $h \in H_\rho$ we can set $\tilde h = h - (\varphi h) u$ so that $\varphi \tilde h = 0$ and thus }
\[ D_{\mu} F_t \tilde h = \tro_{t\varphi \mu} \tilde h = \tro_{t\varphi \mu} h, \]
\Rone{implying the required inclusion of images.}

\bibliographystyle{siam}%{elsarticle-harv}
\bibliography{phd}

\begin{thebibliography}{10}

\bibitem{Baladi14}
{\sc V.~Baladi}, {\em {Linear response, or else}}, in {ICM Seoul 2014,
  Proceedings, Volume III}, Aug 2014, pp.~525--545.

\bibitem{BaladiZetaBook}
{\sc V.~Baladi}, {\em {Dynamical Zeta Functions and Dynamical Determinants for
  Hyperbolic Maps}}, Springer, Berlin, 2016.

\bibitem{Baladi15}
{\sc V.~Baladi, M.~Benedicks, and D.~Schnellmann}, {\em Whitney--{H}{\"o}lder
  continuity of the {SRB} measure for transversal families of smooth unimodal
  maps}, Inventiones mathematicae, 201 (2015), pp.~773--844.

\bibitem{Bandtlow20}
{\sc O.~F. Bandtlow and J.~Slipantschuk}, {\em Lagrange approximation of
  transfer operators associated with holomorphic data}, arXiv preprint
  arXiv:2004.03534,  (2020).

\bibitem{Bell80}
{\sc T.~L. {Bell}}, {\em {Climate sensitivity from fluctuation dissipation:
  Some simple model tests}}, Journal of the Atmospheric Sciences, 37 (1980),
  pp.~1700--1707.

\bibitem{Berger16generic}
{\sc P.~Berger}, {\em Generic family with robustly infinitely many sinks},
  Inventiones mathematicae, 205 (2016), pp.~121--172.

\bibitem{Berger16dimension}
{\sc P.~Berger and J.~De~Simoi}, {\em On the {H}ausdorff dimension of
  {N}ewhouse phenomena}, in Annales Henri Poincar{\'e}, vol.~17, Springer,
  2016, pp.~227--249.

\bibitem{Blumenthal17}
{\sc A.~Blumenthal, J.~Xue, and L.-S. Young}, {\em {Lyapunov exponents for
  random perturbations of some area-preserving maps including the standard
  map}}, Annals of Mathematics,  (2017), pp.~285--310.

\bibitem{Bonatti96}
{\sc C.~Bonatti and L.~J. D{\'\i}az}, {\em Persistent nonhyperbolic transitive
  diffeomorphisms}, Annals of Mathematics, 143 (1996), pp.~357--396.

\bibitem{Bonatti03}
{\sc C.~Bonatti, L.~J. D{\'\i}az, and E.~R. Pujals}, {\em A c1-generic
  dichotomy for diffeomorphisms: weak forms of hyperbolicity or infinitely many
  sinks or sources}, Annals of Mathematics,  (2003), pp.~355--418.

\bibitem{Bowen08}
{\sc R.~Bowen}, {\em Equilibrium states and the ergodic theory of Anosov
  diffeomorphisms}, Lecture notes in mathematics, 470, Springer, Berlin,
  2nd~ed., 2008.

\bibitem{Chazottes05}
{\sc J.-R. Chazottes and B.~Fernandez}, {\em Dynamics of coupled map lattices
  and of related spatially extended systems}, vol.~671, Springer Science \&
  Business Media, 2005.

\bibitem{Chekroun14}
{\sc M.~D. Chekroun, J.~D. Neelin, D.~Kondrashov, J.~C. McWilliams, and
  M.~Ghil}, {\em {Rough parameter dependence in climate models and the role of
  Ruelle-Pollicott resonances}}, Proceedings of the National Academy of
  Sciences, 111 (2014), pp.~1684--90.

\bibitem{Ciszak21}
{\sc M.~Ciszak, S.~Olmi, G.~Innocenti, A.~Torcini, and F.~Marino}, {\em
  Collective canard explosions of globally-coupled rotators with adaptive
  coupling}, 2021.

\bibitem{Cooper13}
{\sc F.~Cooper and P.~Haynes}, {\em {Assessment of the fluctuation-dissipation
  theorem as an estimator of the tropospheric response to forcing}}, Preprint,
  (2013).

\bibitem{Galatolo21}
{\sc S.~Galatolo}, {\em Self consistent transfer operators in a weak coupling
  regime. {I}nvariant measures, convergence to equilibrium, linear reponse and
  control of the statistical properties}, arXiv preprint arXiv:2105.12388,
  (2021).

\bibitem{Gallavotti20}
{\sc G.~Gallavotti}, {\em Nonequilibrium and fluctuation relation}, Journal of
  Statistical Physics, 180 (2020), pp.~172--226.

\bibitem{GallavottiCohen95a}
{\sc G.~Gallavotti and E.~G.~D. Cohen}, {\em Dynamical ensembles in
  nonequilibrium statistical mechanics}, Phys. Rev. Lett., 74 (1995),
  pp.~2694--2697.

\bibitem{GallavottiCohen95b}
\leavevmode\vrule height 2pt depth -1.6pt width 23pt, {\em {{Dy}namical
  ensembles in stationary states}}, Journal of Statistical Physics, 80 (1995),
  pp.~931--970.

\bibitem{Gouezel06}
{\sc S.~Gou{\"e}zel and C.~Liverani}, {\em Banach spaces adapted to anosov
  systems}, Ergodic Theory and dynamical systems, 26 (2006), pp.~189--217.

\bibitem{Gritsun13}
{\sc A.~Gritsun}, {\em Statistical characteristics, circulation regimes and
  unstable periodic orbits of a barotropic atmospheric model}, Philosophical
  Transactions of the Royal Society A: Mathematical, Physical and Engineering
  Sciences, 371 (2013), p.~20120336.

\bibitem{Henry06}
{\sc D.~Henry}, {\em Geometric theory of semilinear parabolic equations},
  vol.~840, Springer, 2006.

\bibitem{Kaneko89}
{\sc K.~Kaneko}, {\em Self-consistent {P}erron-{F}robenius operator for
  spatiotemporal chaos}, Physics Letters A, 139 (1989), pp.~47--52.

\bibitem{Keller00}
{\sc G.~Keller}, {\em An ergodic theoretic approach to mean field coupled
  maps}, in Fractal geometry and stochastics II, Springer, 2000, pp.~183--208.

\bibitem{Koiller10}
{\sc J.~Koiller and L.-S. Young}, {\em Coupled map networks}, Nonlinearity, 23
  (2010), p.~1121.

\bibitem{Kuznetsov21}
{\sc S.~P. Kuznetsov}, {\em Possible occurrence of hyperbolic attractors}, in
  Hyperbolic Chaos, Springer, 2012, pp.~35--56.

\bibitem{Lebowitz99}
{\sc J.~L. Lebowitz and H.~Spohn}, {\em A {G}allavotti--{C}ohen-type symmetry
  in the large deviation functional for stochastic dynamics}, Journal of
  Statistical Physics, 95 (1999), pp.~333--365.

\bibitem{Lembo19}
{\sc V.~Lembo, V.~Lucarini, and F.~Ragone}, {\em Beyond forcing scenarios:
  Predicting climate change through response operators in a coupled general
  circulation model}, arXiv preprint arXiv:1912.03996,  (2019).

\bibitem{Lepri98}
{\sc S.~Lepri, R.~Livi, and A.~Politi}, {\em Energy transport in anharmonic
  lattices close to and far from equilibrium}, Physica D: Nonlinear Phenomena,
  119 (1998), pp.~140--147.

\bibitem{Lucarini14}
{\sc V.~Lucarini, D.~Faranda, J.~Wouters, and T.~Kuna}, {\em Towards a general
  theory of extremes for observables of chaotic dynamical systems}, Journal of
  statistical physics, 154 (2014), pp.~723--750.

\bibitem{Lucarini20}
{\sc V.~Lucarini and A.~Gritsun}, {\em A new mathematical framework for
  atmospheric blocking events}, Climate Dynamics, 54 (2020), pp.~575--598.

\bibitem{Mora93}
{\sc L.~Mora and M.~Viana}, {\em Abundance of strange attractors}, Acta
  mathematica, 171 (1993), pp.~1--71.

\bibitem{Newhouse74}
{\sc S.~E. Newhouse}, {\em Diffeomorphisms with infinitely many sinks},
  Topology, 13 (1974), pp.~9--18.

\bibitem{Newhouse79}
\leavevmode\vrule height 2pt depth -1.6pt width 23pt, {\em The abundance of
  wild hyperbolic sets and non-smooth stable sets for diffeomorphisms},
  Publications Math{\'e}matiques de l'IH{\'E}S, 50 (1979), pp.~101--151.

\bibitem{ApproxFun}
{\sc S.~Olver}, {\em {Approx{F}un}}, 2019.
\newblock Available at \url{https://github.com/JuliaApproximation/ApproxFun.jl}
  and in the Julia package repository.

\bibitem{Palis08}
{\sc J.~Palis}, {\em Open questions leading to a global perspective in
  dynamics}, Nonlinearity, 21 (2008), pp.~1--37.

\bibitem{Palis94}
{\sc J.~Palis and M.~Viana}, {\em High dimension diffeomorphisms displaying
  infinitely many periodic attractors}, Annals of mathematics,  (1994),
  pp.~207--250.

\bibitem{Pikovsky94}
{\sc A.~S. Pikovsky and J.~Kurths}, {\em {Do globally coupled maps really
  violate the law of large numbers?}}, Phys. Rev. Lett., 72 (1994),
  pp.~1644--1646.

\bibitem{RagoneEtAl15}
{\sc F.~{Ragone}, V.~{Lucarini}, and F.~{Lunkeit}}, {\em {A new framework for
  climate sensitivity and prediction: a modelling perspective}}, Climate
  Dynamics, 46 (2016), pp.~1459--1471.

\bibitem{Ruelle97}
{\sc D.~Ruelle}, {\em {Differentiation of {SRB} states}}, Communications in
  Mathematical Physics, 187 (1997), pp.~227--241.

\bibitem{Ruelle18}
\leavevmode\vrule height 2pt depth -1.6pt width 23pt, {\em {Linear response
  theory for diffeomorphisms with tangencies of stable and unstable
  manifolds---a contribution to the {G}allavotti-{C}ohen chaotic hypothesis}},
  Nonlinearity, 31 (2018), p.~5683.

\bibitem{Sander00}
{\sc E.~Sander}, {\em Homoclinic tangles for noninvertible maps}, Nonlinear
  Analysis: Theory, Methods \& Applications, 41 (2000), pp.~259--276.

\bibitem{Sedro18}
{\sc J.~Sedro}, {\em A regularity result for fixed points, with applications to
  linear response}, Nonlinearity, 31 (2018), p.~1417.

\bibitem{Selley19}
{\sc F.~M. S{\'e}lley}, {\em A self-consistent dynamical system with multiple
  absolutely continuous invariant measures}, arXiv preprint arXiv:1909.04484,
  (2019).

\bibitem{Selley21}
{\sc F.~M. S{\'e}lley and M.~Tanzi}, {\em Linear response for a family of
  self-consistent transfer operators}, Communications in Mathematical Physics,
  382 (2021), pp.~1601--1624.

\bibitem{Tantet18}
{\sc A.~Tantet, V.~Lucarini, F.~Lunkeit, and H.~A. Dijkstra}, {\em Crisis of
  the chaotic attractor of a climate model: a transfer operator approach},
  Nonlinearity, 31 (2018), p.~2221.

\bibitem{Trefethen13}
{\sc L.~N. Trefethen}, {\em {Approximation theory and approximation practice}},
  Siam, Philadelphia, PA, 2013.

\bibitem{VanVliet08}
{\sc C.~M. Van~Vliet}, {\em Equilibrium And Non-equilibrium Statistical
  Mechanics}, World Scientific Publishing Company, 2008.

\bibitem{Poltergeist}
{\sc C.~L. Wormell}, {\em {Poltergeist}}, 2017.
\newblock Available at \url{https://github.com/wormell/Poltergeist.jl} and in
  the Julia package repository.

\bibitem{Wormell19}
\leavevmode\vrule height 2pt depth -1.6pt width 23pt, {\em {Spectral {G}alerkin
  methods for transfer operators in uniformly expanding dynamics}}, Numerische
  Mathematik, 142 (2019), pp.~421--463.

\bibitem{WormellGottwald18}
{\sc C.~L. Wormell and G.~A. Gottwald}, {\em On the validity of linear response
  theory in high-dimensional deterministic dynamical systems}, Journal of
  Statistical Physics, 172 (2018), pp.~1479--1498.

\bibitem{WormellGottwald19}
{\sc C.~L. Wormell and G.~A. Gottwald}, {\em {Linear response for macroscopic
  observables in high-dimensional systems}}, Chaos: An Interdisciplinary
  Journal of Nonlinear Science, 29 (2019), p.~113127.

\bibitem{Yorke83}
{\sc J.~A. Yorke and K.~T. Alligood}, {\em Cascades of period-doubling
  bifurcations: a prerequisite for horseshoes}, Bulletin (New Series) of the
  American Mathematical Society, 9 (1983), pp.~319--322.

\bibitem{Young95}
{\sc L.-S. Young}, {\em Ergodic theory of differentiable dynamical systems}, in
  Real and complex dynamical systems, Springer, 1995, pp.~293--336.

\bibitem{Young19}
\leavevmode\vrule height 2pt depth -1.6pt width 23pt, {\em Comparing chaotic
  and random dynamical systems}, Journal of Mathematical Physics, 60 (2019),
  p.~052701.

\end{thebibliography}

\end{document}